\documentstyle[12pt,aasms4]{article}
\begin{document}   
\title{Dynamic Screening in Thermonuclear Reactions}
\author{Merav Opher and Reuven Opher}

\affil{Instituto Astron\^omico e Geof\'\i sico, Universidade de 
S\~ao Paulo\\%
Av. Miguel St\'efano 4200, S\~ao Paulo, 04301-904, SP, Brazil}
\authoremail{merav@orion.iagusp.usp.br}

\abstract{It has recently been argued that there are no dynamic 
screening corrections to Salpeter's 
enhancement factor in thermonuclear reactions, 
in the weak-screening limit. The arguments 
used were: 1) The Gibbs probability distribution is factorable into 
two parts, one of which, $exp(-\beta \sum e_{i}e_{j}/r_{ij})$ 
($\beta=1/k_{B}T$), is independent of velocity space; and 2) The 
enhancement factor 
is $w=1+\beta^{2}e^{2}Z_{1}Z_{2}\langle \phi^{2} \rangle$ with 
${\langle \phi^{2} \rangle}_{k}={\langle E^{2} \rangle}_{k}/k^{2}$ 
and ${\langle E^{2} \rangle}_{k}/(8\pi)=(T/2)[1-\varepsilon^{-1}
(0,k)]$. We show that both of these arguments are incorrect.}

\section{Introduction}
Our knowledge of reaction rates in the Sun is becoming more and more
accurate. There have been several works dealing with electrostatic 
screening effects in the solar plasma (as for example, 
Carraro, Sch\"{a}fer, \& Koonin 1988; Gruzinov \& Bahcall 1997; 
Br\"{u}ggen \& Gough 1997). Recently, more precise calculations have been 
made beyond the linear regime (Gruzinov \& Bahcall 1998). 

The screening has the effect of lowering the Coulomb barrier between 
the interacting ions, therefore enhancing the reaction rates. This is 
included in the enhancing factor in the weak screening 
limit (where $Z_{1}Z_{2}e^{2}/R_{D}T \ll 1$ and $R_{D}$ is 
the Debye radius) 
\begin{equation}
w=exp~\lambda~,
\end{equation}
where
\begin{equation}
\lambda=Z_{1}Z_{2}e^{2}/R_{D}T~.
\end{equation}
This screeening factor uses the Debye-H\"{u}ckel expression. 

Usually, the calculation is made in the electrostatic case, based on the 
calculation of Salpeter (1954). It is assumed that the motion of the screened ion 
is slow, compared to the motion of the screening particles. 
Carraro, Sch\"{a}fer, \& Koonin (1988) have studied the case of dynamic screening. 

In a recent study, Gruzinov (1998) argued that in the weak screening limit, 
there is no dynamic screening corrections to Salpeter's enhancement 
factor, even for high energies. He based his conclusion on two arguments: 
1) For thermodynamic equilibrium, the Gibbs probability distributions 
in velocity and configuration space are decoupled; and 
2) Through an analysis of the thermal electric field, we can 
estimate the random electrostatic potential and conclude that the 
enhancement of the reaction rates is given by the 
Salpeter expression. 

We show below that although these arguments are simple, 
a careful examination shows that they are wrong. 
In \S 2 and \S 3 we recall the arguments and show why 
they are incorrect. 

\section{Gibbs Distribution}
The enhancement factor for a reaction rate between nuclei of charges 
$Z_{1}e$ and $Z_{2}e$ is 
\begin{equation}
w=exp(-Z_{2}e\phi_{0}/T)
\end{equation}
($k_{B}=1$), where $\phi_{0}$ is the electrostatic potential, created by 
the plasma on the particle $Z_{1}e$. 

A test particle $Z_{1}e$ moving through a plasma with velocity $v'$ suffers 
dynamic screening. The electrostatic potential is written as (see 
Krall \& Trivelpiece (1973), chap.11)
\begin{equation}
\phi_{0}=4\pi e Z_{1} \int \frac{d^{3}k}{{(2\pi)}^{3}}
\left [ \frac{1}{\varepsilon(k,kv')}-1 \right ] k^{-2}~,
\end{equation}
where $\varepsilon$ is the dielectric permittivity, which describes the plasma 
response to the test particle. ($\varepsilon$ depends on the velocity 
distribution of the plasma particles $f(v)$, which usually is taken as Maxwellian). 
From the above expression, it can be seen that $\phi_{0}$ depends on the 
velocity $v'$ of particle $Z_{1}e$. Gruzinov (1998) argued that the 
Gibbs probability distribution $\rho$ is
\begin{equation}
\rho \sim exp \left ( -\beta \sum \frac{m_{i}v_{i}^{2}}{2} -\beta 
\sum \frac{e_{i}e_{j}}{r_{ij}} \right ) ~,
\end{equation}
(where $\beta=1/T$) which can be factorable into
\begin{equation}
\rho \sim exp \left ( -\beta \sum \frac{m_{i}v_{i}^{2}}{2} \right ) 
exp \left ( -\beta \sum \frac{e_{i}e_{j}}{r_{ij}} \right ) ~.
\end{equation}
From the above, it was argued that the distributions in velocity and 
configuration space are decoupled. 

This first argument is simple. However, it is based on a 
misunderstanding. The general Gibbs probability distribution $\rho$ 
for a plasma is
\begin{equation}
\rho \sim exp \left ( -\beta \sum \frac{m_{i}v_{i}^{2}}{2} \right ) 
exp \left ( -\beta \sum_{i > k} \sum_{k} W_{ik} \right ) ~,
\end{equation}
where $W_{ik}$ is the interaction energy between the particles. 
$W_{ik}$ is the interaction energy of particle $i$ with all the other particles 
in the plasma. This energy is the Coulomb energy 
${e_{i}e_{j}}/{r_{ij}}$, related to the positions 
of all the other particles. It is an assumption that 
their positions $r_{i}$, $r_{j}$ are independent of their velocities.
However, we know from dynamic screening, Eq. (4), that their coordinates 
are, in fact, velocity dependent. Eq. (6) states that the coordinates are 
independent of the velocities only in zero order. The exact Gibbs 
distribution takes into account dynamic corrections. 

In fact, this argument is circular. The separability of the Gibbs 
distribution in velocity and configuration contributions 
is true only if the particles are moving 
under a (static) conservative force. The argument to show that there 
are no dynamical contributions, therefore, rests on a statement that is 
valid only if there are no dynamical contributions, which, in fact, is what 
he set out to prove.

\section{Thermal Electric Field}
Due to thermal fluctuations in the plasma, the reaction rate between two 
fast moving ions $Z_{1}e$ and $Z_{2}e$ is enhanced. The enhancement factor 
of a reaction rate is 
\begin{equation}
w=1+ \beta^{2} e^{2}Z_{1}Z_{2} \langle \phi^{2} \rangle  ~,
\end{equation}
where $\langle \phi^{2} \rangle$ is the average of the square 
of the random electrostatic potential $\phi$. $\langle \phi^{2} \rangle$ is 
given by 
\begin{equation}
\langle \phi^{2} \rangle=\int \frac{d^{3}k}{{(2\pi)}^{3}} 
{\langle \phi^{2} \rangle}_{k}~,
\end{equation}
where ${\langle \phi^{2} \rangle}_{k}={\langle E^{2} \rangle}_{k}/k^{2}$. 

In his second argument, the expression used for the fluctuation electric 
field was (Krall \& Trivelpice (1973), chap.11)
\begin{equation}
\frac{{\langle E^{2} \rangle}_{k}}{(8\pi)}=
\left ( \frac{T}{2} \right ) 
\left [ 1- \frac{1}{\varepsilon(0,k)} \right ] ~.
\end{equation}
Substituing this expression in Eq.(8) and Eq.(9), 
Salpeter's expression is obtained. 

However, the expression of the thermal electric field (for 
a Maxwellian plasma), given by Eq. (10), assumes that $\omega \ll T$ 
($\hbar, k_{B}=1$). The general expression for the intensity of the 
electric field is given by the {\it Fluctuation-Dissipation Theorem} 
(see for example Sitenko (1967), Akhiezer et al. (1975)):
\begin{equation}
\frac{{\langle E^{2} \rangle}_{k\omega}}{(8\pi)}
=\frac{1}{e^{{\omega}/T}-1}\frac{Im \varepsilon}
{{\mid \varepsilon \mid}^{2}}
\end{equation}
and
\begin{equation}
\frac{{\langle E^{2} \rangle}_{k}}{(8\pi)}
=\int d\omega \frac{1}{e^{{\omega}/T}-1}\frac{Im \varepsilon}
{{\mid \varepsilon \mid}^{2}}~.
\end{equation}
This expression includes electrostatic Langmuir waves as well as all 
other fluctuations that exist in a plasma. 
In the limit of $\omega \ll T$, 
\begin{equation}
\frac{{\langle E^{2} \rangle}_{k}}{(8\pi)}
=\int d\omega \frac{T}{\omega}\frac{Im \varepsilon}
{{\mid \varepsilon \mid}^{2}}~,
\end{equation}
for which it is then possible to use the Kramers-Kronig relations. Eq.(13) then 
turns out to be
\begin{equation}
\frac{{\langle E^{2} \rangle}_{k}}{(8\pi)}=
\left ( \frac{T}{2} \right ) 
\left [ 1- \frac{1}{\varepsilon(0,k)} \right ] ~,
\end{equation}
which is the expression used. 

The assumption that $\omega \ll T$, however, is very strong. In fact, 
in the case of the transverse electric field, we showed 
(Opher \& Opher (1997a, 1997b)) that only by not making this strong 
assumption, is the blackbody at high frequencies obtained. 
We recently showed (Opher \& Opher (1999)), that by not making 
the assumption that $\omega \ll T$, the energy of a plasma in the classical 
limit is larger than previously thought.

Without assuming that $\omega \ll T$, the enhancement factor is 
given by Eq. (8) with Eq. (12). 

It is also to be noted that the second argument is also circular. It assumes 
that $\omega \ll T$, making it a static analysis, which is then used to 
prove that there does not exist a dynamic contribution.

It is to be emphasized that we are not arguing here whether or not 
dynamic screening exists in thermonuclear reactions. 
We only show that the arguments used by Gruzinov (1998), to prove 
that dynamic screening does not exist, are not valid.

\acknowledgments 
M.O. would like to thank the Brazilian agency FAPESP
for support (no. 97/13427-8). R.O. thanks the Brazilian agency CNPq for 
partial support. Both authors would also like to thank the Brazilian project 
Pronex/FINEP (no. 41.96.0908.00) for support.


\begin{thebibliography}{99}

\bibitem{r1} Akhiezer, A. I., Akhiezer, I. A., Plovin, R. V., 
Sitenko, A. G., \& Stepanov, K. N. 1975, Plasma Electrodynamics, Vol. 2 
(Pergamon Press, Oxford).
\bibitem{r11} Brown, L. S., \& Sawyer, R. F. 1997 \apj, 489, 968.
\bibitem{r2} Br\"{u}ggen, M., \& Gough, D. O. 1997, \apj, 488, 867.
\bibitem{r3} Carraro, C., Sch\"{a}fer, A., \& Koonin, S. E. 1988, 
\apj, 331, 565.
\bibitem{r4} Gruzinov, A. V. 1998, \apj, 496, 503.
\bibitem{r6} Gruzinov, A. V., \& Bahcall, J. N. 1998 \apj 504, 996.
\bibitem{r5} Gruzinov, A. V., \& Bahcall, J. N. 1997 \apj, 490, 437.
\bibitem{r6} Opher, M., \& Opher, R. 1997, \prl, 79, 2628.
\bibitem{r7} Opher, M., \& Opher, R. 1997, \prd, 56, 3296.
\bibitem{r8} Opher, M. \& Opher, R. 1999, \prl, 82, 4835.
\bibitem{r9} Krall, N., \& Trivelpiece, A. W. 1973, 
Principles of Plasma Physics (New York: McGraw-Hill).
\bibitem{r10} Salpeter, E. E. 1954, Australian J.Phys., 7, 373.
\bibitem{r11} Sitenko, A. G. 1967, Electromagnetic Fluctuations in Plasma
(Academic Press, NY).
\end{thebibliography}
\end{document}